\documentclass[12pt]{article}

\usepackage{sbc-template}

\usepackage{graphicx,url}

\usepackage{pdflscape}
\usepackage{scalefnt}
\usepackage{multirow}

\usepackage[brazil]{babel}   
\usepackage[utf8]{inputenc}

\title{Requirements Engineering, Software Testing and Education: A Systematic Mapping}

\author{Thalia S. Santana\inst{1}, Taciana N. Kudo\inst{1}, Renato F. Bulcão-Neto\inst{1}}

\address{Instituto de Informática -- Universidade Federal de Goiás (UFG)\\
    Caixa Postal 131 – 74.690-900 – Goiânia – GO – Brasil
\email{thaliasantana@discente.ufg.br, \{taciana,rbulcao\}@ufg.br}
}

\begin{document} 

\maketitle

\begin{abstract}
The activities of requirements engineering and software testing are intrinsically related to each other, as these two areas are linked when seeking to specify and also ensure the expectations of a software product, with quality and on time. This systematic mapping study aims to verify how requirements and testing are being addressed together in the educational context.
\end{abstract}
     
\begin{resumo} 
As atividades de engenharia de requisitos e de teste de software possuem relação intrínseca entre si, ao passo que estas duas áreas encontram-se vinculadas ao buscar especificar e também garantir as expectativas de um produto de software, com qualidade e no prazo adequado. O presente mapeamento sistemático da literatura visa verificar como requisitos e testes estão sendo abordados de forma conjunta no contexto educacional.
\end{resumo}

\section{Introdução}

Dentro da Engenharia de Requisitos (ER), requisitos de software devem representar funcionalidades e comportamentos que se esperam de um sistema, enquanto  Testes de Software (TS) visam atestar se os requisitos foram corretamente atendidos. Devido a essa sinergia, é essencial que atividades de ER e TS possam estar alinhadas diretamente \cite{barmi2011}. Assim, requisitos e testes deveriam estar vinculados em prol de evitar possíveis problemas na entrega de produtos de software \cite{coutinho2019}.

Como apontado pelos estudos de \cite{barmi2011, bjarnason2014, unterkalmsteiner2015, bjarnason2016}, o alinhamento de requisitos e testes traz benefícios para a indústria de software. Para que isso possa ser promovido no mercado de trabalho, é necessário investigar a formação profissional, em vistas de compreender se estão sendo trabalhadas habilidades necessárias para interligar tais atividades, por meio do modo em que ER e TS são ensinados nas universidades. De acordo com \cite{fernandez2017}, a falta de qualificação profissional vêm impactando no êxito de projetos de software -- o que demonstra a importância de averiguar a existência de trabalhos que adotem o alinhamento destas duas atividades no âmbito educacional.

Em prol de auxiliar na definição de temas de pesquisa de mestrado ou doutorado, o Mapeamento Sistemático da Literatura (MSL) vêm sendo empregado como uma forma ampla e repetível de investigação em tópicos de estudo específicos. Dentre seus principais benefícios, cita-se o entendimento da extensão da pesquisa, a sumarização de achados derivados de estudos publicados em uma determinada área de conhecimento e a detecção de lacunas de pesquisa promissoras em um campo de estudo \cite{nakagawa2017}.

Diante disso, dada a existência de MSLs que verificaram isoladamente o estado da arte do ensino tanto de ER \cite{ouhbi2015} quanto de TS \cite{garousi2020msl}, o presente MSL diferencia-se ao buscar compreender a intersecção de ambas as áreas na educação. Portanto, o objetivo é apresentar informações quanto ao estado atual de pesquisas relacionadas ao ensino de ER e TS.  

Sendo assim, a Seção \ref{sec:msl} detalha as atividades desenvolvidas para a realização deste MSL, a saber: questões de pesquisa e termos de busca; \textit{string} de pesquisa e estratégia de busca adotada; seleção dos estudos e extração de dados. Por conseguinte, a Seção \ref{sec:sintese} traz os principais apontamentos e resultados decorrentes do MSL, seguido da considerações finais e trabalhos futuros na Seção \ref{sec:consid_msl}.

\section{Mapeamento Sistemático}
\label{sec:msl}

Um MSL é um tipo de estudo secundário com o objetivo de prover uma visão geral de uma área de pesquisa, por meio da categorização e classificação de resultados advindos da literatura \cite{petersen2015}. Para tanto, deve adotar um processo sistematizado para a identificação, análise e interpretação de evidências acerca de uma determinada temática, permitindo sua reprodutibilidade \cite{nakagawa2017}. 

Nesse sentido, o presente trabalho objetivou identificar o estado da arte referente ao alinhamento de ER e TS no ensino. Idealmente, são estudos-alvo deste mapeamento os estudos primários que descreverem estudos de caso, experimentos e relatos de experiência com algum tipo de avaliação ou validação da pesquisa, publicados em periódicos e anais de eventos científicos em um recorte temporal de cinco anos (2017 -- 2021).

Estudos de MSL incluem três etapas de execução: planejamento, condução e publicação de resultados. Sendo assim, a plataforma \textit{web Parsif.al}\footnote{https://parsif.al/} foi utilizada como ferramenta de apoio para o MSL, em especial para atividades de planejamento e condução -- desde a elaboração de um protocolo de estudos sistemáticos até a extração de dados. Em consonância, o processo de análise e síntese dos dados obtidos foi efetuado por intermédio de planilhas do \textit{Google Sheets}\footnote{https://sheets.google.com/}, incluindo a geração de gráficos e visualizações.

\subsection{Questões de Pesquisa e Termos de Busca}

Com o intuito de mapear o estado da arte do ensino de ER e TS, foi elaborado um protocolo sistematizado para a realização do MSL. Assim, definiu-se uma questão de pesquisa (QP) para nortear este MSL. Portanto, a QP1 \textit{busca entender qual é o estado da arte sobre o ensino de ER e de TS}.

Visando responder a esse questionamento, o próximo passo foi a identificação de estudos relevantes para responder a QP estabelecida. Inicialmente, foram realizados testes-piloto com palavras-chave que remetessem as atividades de requisitos e testes presentes no ciclo de desenvolvimento de software. De forma conjunta, também foram adotados sinônimos relativos ao ensino. Todavia, em primeira instância, percebeu-se que muitos trabalhos não relevantes estavam sendo retornados. 

Em prol de obter o balanceamento entre a precisão e a recuperação dos resultados, com o auxílio de especialistas em ER, foi constatada a existência de terminologias específicas para tratar de educação em ER e TS, sendo: \textit{Requirements Engineering Education} (em português, EER -- Educação em Engenharia de Requisitos) e \textit{Software Testing Education} (em português, Educação em Teste de Software). Ambos os termos foram apresentados em dois MSLs específicos de educação em cada uma das áreas  \cite{ouhbi2015, garousi2020msl}.

\subsection{String e Busca Automática}

O equilíbrio das terminologias averiguadas nos testes-piloto deram origem à \textit{string} de busca final. A mesma foi aplicada nos mecanismos de busca considerando sua abrangência no título, resumo e palavras-chave, como apresentada a seguir:

\begin{center}
\textit{\textbf{(``requirements engineering education'' OR ``software testing education'')}}
\end{center}

Por meio de uma estratégia de busca automática, as bases de dados e motores de busca selecionados trataram-se de \textit{ACM Digital Library}\footnote{https://dl.acm.org/}, \textit{IEEE Xplore}\footnote{https://ieeexplore.ieee.org/}, \textit{Scopus}\footnote{https://www.scopus.com/} e \textit{SBC Open Lib}\footnote{https://sol.sbc.org.br/} -- escolhidas por representarem algumas das principais bases bibliográficas em Ciência da Computação \cite{nakagawa2017}. Não foram utilizadas demais estratégias de busca combinadas.

\subsection{Seleção dos Estudos e Extração de Dados}

Para decidir quais estudos seriam selecionados no MSL, critérios de inclusão (CI) e exclusão (CE) foram definidos. Um único CI determina se o trabalho é aceito, sendo incluído ao \textit{apresentar uma contribuição no ensino de ER ou de TS ou ambos}. Quanto aos CE, é necessário atender a pelo menos um dos CE definidos para eliminação do trabalho. Desta maneira, utilizou-se nove itens para a exclusão dos estudos, a seguir:

\begin{itemize}
    \item CE1 -- O texto completo não se encontra disponível;
    \item CE2 -- O texto completo não está escrito em inglês, português ou espanhol;
    \item CE3 -- O trabalho não é um artigo completo de conferência ou periódico;
    \item CE4 -- O trabalho não é um estudo primário;
    \item CE5 -- O estudo é uma versão mais antiga de outro já considerado;
    \item CE6 -- O artigo foi publicado antes de 2017;
    \item CE7 -- O estudo não trata de ensino em engenharia de software;
    \item CE8 -- O estudo não contempla as atividades de engenharia de requisitos e/ou teste de software;
    \item CE9 -- O estudo não envolve uma pesquisa de avaliação, validação ou relato de experiência.
\end{itemize}

\begin{figure}[ht]
  \centering
  \includegraphics[width=\textwidth]{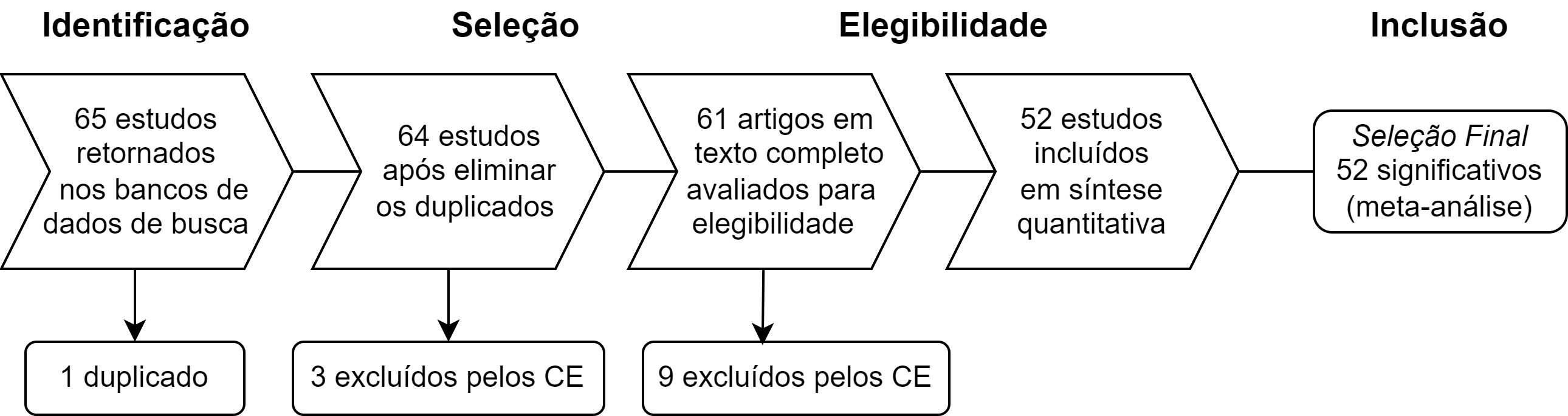}
  \caption{Fluxo de informação com as etapas e quantitativo de estudos aceitos.}
  \label{fig_estudos}
\end{figure}

Na condução do MSL, após a recuperação dos resultados advindos da aplicação da \textit{string}\footnote{A busca em todas as bibliotecas digitais deu-se em 02/03/2022.} nos mecanismos de busca, 65 trabalhos foram retornados dentro do filtro temporal de cinco anos completos. Destes, apenas um estudo foi duplicado. Após a leitura dos metadados e posteriormente, do texto completo, 52 estudos foram considerados como relevantes para este MSL (após a devida aplicação dos CI e CE). A Figura \ref{fig_estudos} ilustra a quantidade de
estudos identificados, duplicados e selecionados neste MSL.

A lista completa contendo os 52 trabalhos por título, fonte de busca e referência encontra-se disponível na Tabela \ref{tab:artigos_aceitos}.

\begin{table}[!htbp]
\centering
\scalefont{0.8}
\caption{Trabalhos relevantes analisados neste MSL.}
	\label{tab:artigos_aceitos}
\begin{tabular}{lll}
\hline
\textbf{Título} & \textbf{Fonte} & \textbf{Referência}\\ \hline
\textit{\begin{tabular}[c]{@{}l@{}} A Case Study on the Application of Case-Based \\ Learning in Software Testing\end{tabular}} & \textit{ACM DL} & \cite{tiwari2018} \\ \hline

\textit{\begin{tabular}[c]{@{}l@{}} A Chatterbot Sensitive to Student's Context to Help on\\ Software Engineering Education\end{tabular}} & \textit{IEEE Xplore} & \cite{paschoal2018} \\ \hline

\textit{\begin{tabular}[c]{@{}l@{}} A Comparison of Inquiry-Based Conceptual Feedback\\ vs. Traditional Detailed Feedback Mechanisms\\ in Software Testing Education: An Empirical\\ Investigation\end{tabular}} & \textit{Scopus} & \cite{cordova2021} \\ \hline

\textit{\begin{tabular}[c]{@{}l@{}} A diagnosis on software testing education in the\\ Brazilian Universities\end{tabular}} & \textit{IEEE Xplore} & \cite{elgrably2021} \\ \hline

\textit{\begin{tabular}[c]{@{}l@{}} A Gamified Tutorial for Learning About Security\\ Requirements Engineering\end{tabular}} & \textit{IEEE Xplore} & \cite{alami2017} \\ \hline

\textit{\begin{tabular}[c]{@{}l@{}} A qualitative study of teaching requirements\\ engineering in universities\end{tabular}} & \textit{Scopus} & \cite{epifanio2019} \\ \hline

\textit{\begin{tabular}[c]{@{}l@{}} A survey on graduates' curriculum-based knowledge\\ gaps in software testing\end{tabular}} & \textit{Scopus} & \cite{scatalon2019} \\ \hline

\textit{\begin{tabular}[c]{@{}l@{}} A survey on software testing education in Brazil\end{tabular}} & \textit{Scopus} & \cite{paschoal2018b}\\ \hline

\textit{\begin{tabular}[c]{@{}l@{}} An experimental evaluation of peer testing in the\\ context of the teaching of software testing\end{tabular}} & \textit{IEEE Xplore} & \cite{barbosa2017} \\ \hline

\textit{\begin{tabular}[c]{@{}l@{}} An Undergraduate Requirements Engineering \\Curriculum with Formal Methods\end{tabular}} & \textit{IEEE Xplore} & \cite{westphal2018} \\ \hline

\textit{\begin{tabular}[c]{@{}l@{}} Analyzing Competences in Software Testing:\\
Combining Thematic Analysis with Natural Language\\
Processing (NLP)\end{tabular}} & \textit{IEEE Xplore} & \cite{rahman2021} \\ \hline

\textit{\begin{tabular}[c]{@{}l@{}} Appreciate the journey not the destination - \\Using video assignments in software testing education\end{tabular}} & \textit{Scopus} & \cite{causevic2018} \\ \hline

\textit{\begin{tabular}[c]{@{}l@{}} As a Teacher, I Want to Know What to Teach in\\ Requirements Engineering so That Professionals Can\\
Be Better Prepared\end{tabular}} & \textit{ACM DL} & \cite{benitti2017} \\ \hline

\textit{\begin{tabular}[c]{@{}l@{}} Can scoring rubrics be used in assessing the\\ performance of students in software requirements\\ engineering education?\end{tabular}} & \textit{Scopus} & \cite{mkpojiogu2017} \\ \hline

\textit{\begin{tabular}[c]{@{}l@{}}Chatbot-based Interview Simulator: A Feasible\\ Approach to Train Novice Requirements Engineers\end{tabular}} & \textit{IEEE Xplore} & \cite{laiq2020} \\ \hline

\textit{\begin{tabular}[c]{@{}l@{}}Common mistakes of student analysts in \\requirements elicitation interviews\end{tabular}} & \textit{Scopus} & \cite{donati2017}\\ \hline

\textit{\begin{tabular}[c]{@{}l@{}}Design and Development of a Serious Game for the\\ Teaching of Requirements Elicitation and Analysis\end{tabular}} & \textit{IEEE Xplore} & \cite{ibrahim2019} \\ \hline

\textit{\begin{tabular}[c]{@{}l@{}}Do we preach what we practice? Investigating the\\ practical relevance of requirements engineering\\ syllabi – the IREB case\end{tabular}} & \textit{Scopus} & \cite{fernandez2019} \\ \hline

\textit{\begin{tabular}[c]{@{}l@{}}Educational games: A contribution to software\\ testing education\end{tabular}} & \textit{IEEE Xplore} & \cite{valle2017} \\ \hline

\textit{\begin{tabular}[c]{@{}l@{}}Evaluating Role Playing Efficiency to Teach\\ Requirements Engineering\end{tabular}} & \textit{IEEE Xplore} & \cite{ouhbi2019}\\ \hline

\textit{\begin{tabular}[c]{@{}l@{}}Evaluating the impact of Software Testing\\ Education through the Flipped Classroom Model\\ in deriving Test Requirements\end{tabular}} & \textit{Scopus} & \cite{paschoal2020} \\ \hline

\textit{\begin{tabular}[c]{@{}l@{}}Evaluating the Students' Experience with a\\ requirements elicitation and communication game\end{tabular}} & \textit{Scopus} & \cite{vilela2020}\\ \hline

\textit{\begin{tabular}[c]{@{}l@{}}Experiences in Teaching and Learning Requirements\\ Engineering on a Sound Didactical Basis\end{tabular}} & \textit{ACM DL} & \cite{sedelmaier2017} \\ \hline

\textit{\begin{tabular}[c]{@{}l@{}}From blackboard to the office: A look into how\\ practitioners perceive software testing education\end{tabular}} & \textit{Scopus} & \cite{martins2021} \\ \hline

\textit{\begin{tabular}[c]{@{}l@{}}Gamifying a software testing course with code\\ defenders\end{tabular}} & \textit{Scopus} & \cite{fraser2019} \\ \hline

\end{tabular}
\end{table}

\addtocounter{table}{-1}

\begin{table}[!htbp]
\centering
\scalefont{0.8}
\caption{Trabalhos aceitos após critérios do MSL (\emph{Continuação}).}
\begin{tabular}{lll}
\hline
\textbf{Título} & \textbf{Fonte} & \textbf{Referência} \\ \hline

\textit{\begin{tabular}[c]{@{}l@{}}Good Bug Hunting: Inspiring and Motivating\\ Software Testing Novices\end{tabular}} & \textit{Scopus} & \cite{silvis2021} \\ \hline

\textit{\begin{tabular}[c]{@{}l@{}}Guidelines for software testing education objectives\\ from industry practices with a constructive\\ alignment approach\end{tabular}} & \textit{Scopus} & \cite{hynninen2018} \\ \hline

\textit{\begin{tabular}[c]{@{}l@{}}How Students Unit Test: Perceptions, Practices,\\ and Pitfalls\end{tabular}} & \textit{ACM DL} & \cite{bai2021} \\ \hline

\textit{\begin{tabular}[c]{@{}l@{}}Improving Software Testing Education via Industry\\ Sponsored Contests\end{tabular}} & \textit{IEEE Xplore} & \cite{wong2018} \\ \hline

\textit{\begin{tabular}[c]{@{}l@{}}Inspectors Academy: Pedagogical Design for\\ Requirements Inspection Training\end{tabular}} & \textit{IEEE Xplore} & \cite{bano2020} \\ \hline

\textit{\begin{tabular}[c]{@{}l@{}}Integrating Testing Throughout the CS Curriculum\end{tabular}} & \textit{IEEE Xplore} & \cite{heckman2020} \\ \hline

\textit{\begin{tabular}[c]{@{}l@{}}Involving Customers in Requirements Engineering\\ Education: Mind the Goals!\end{tabular}} & \textit{ACM DL} & \cite{hagel2018} \\ \hline

\textit{\begin{tabular}[c]{@{}l@{}}Is It Worth Using Gamification on Software Testing\\ Education? An Extended Experience Report in\\ the Context of Undergraduate Students\end{tabular}} & \textit{SBC Open Lib} & \cite{jesus2020}\\ \hline

\textit{\begin{tabular}[c]{@{}l@{}}Is Role Playing in Requirements Engineering\\ Education Increasing Learning Outcome?\end{tabular}} & \textit{ACM DL} & \cite{svensson2017} \\ \hline

\textit{\begin{tabular}[c]{@{}l@{}}Mutation Testing and Self/Peer Assessment: \\Analyzing their Effect on Students in a Software \\Testing Course\end{tabular}} & \textit{IEEE Xplore} & \cite{delgado2021}\\ \hline

\textit{\begin{tabular}[c]{@{}l@{}}Pragmatic sotware testing education\end{tabular}} & \textit{Scopus} & \cite{aniche2019} \\ \hline

\textit{\begin{tabular}[c]{@{}l@{}}Requirements Analysis Skills: How to Train \\Practitioners?\end{tabular}} & \textit{IEEE Xplore} & \cite{morales2018} \\ \hline

\textit{\begin{tabular}[c]{@{}l@{}}Requirements Engineering Out of the Classroom:\\ Anticipating Challenges Experienced in Practice\end{tabular}} & \textit{IEEE Xplore} & \cite{marques2020} \\ \hline

\textit{\begin{tabular}[c]{@{}l@{}}SaPeer and ReverseSaPeer: teaching requirements\\ elicitation interviews with role-playing and role\\ reversal\end{tabular}} & \textit{Scopus} & \cite{ferrari2020} \\ \hline

\textit{\begin{tabular}[c]{@{}l@{}}Software testing education through a \\collaborative virtual approach\end{tabular}} & \textit{Scopus} & \cite{ucan2018} \\ \hline

\textit{\begin{tabular}[c]{@{}l@{}}Software testing education: Dreams and challenges\\ when bringing academia and industry \\closer together\end{tabular}} & \textit{Scopus} & \cite{andrade2019} \\ \hline

\textit{\begin{tabular}[c]{@{}l@{}}Software visual specification for requirement\\ engineering education\end{tabular}} & \textit{Scopus} & \cite{zainuddin2019} \\ \hline

\textit{\begin{tabular}[c]{@{}l@{}}Systematic evolution of a learning setting for\\ requirements engineering education based on\\ competence-oriented didactics\end{tabular}} & \textit{IEEE Xplore} & \cite{sedelmaier2018} \\ \hline

\textit{\begin{tabular}[c]{@{}l@{}}Teaching Motivational Models in Agile\\ Requirements Engineering\end{tabular}} & \textit{IEEE Xplore} & \cite{lorca2018} \\ \hline

\textit{\begin{tabular}[c]{@{}l@{}}Teaching requirements elicitation interviews: \\an empirical study of learning from mistakes\end{tabular}} & \textit{Scopus} & \cite{bano2019} \\ \hline

\textit{\begin{tabular}[c]{@{}l@{}}Teaching Software Process Models to Software\\ Engineering Students: An Exploratory Study\end{tabular}} & \textit{IEEE Xplore} & \cite{tiwari2019} \\ \hline

\textit{\begin{tabular}[c]{@{}l@{}}Teaching Software Testing in an Algorithms\\ and Data Structures Course\end{tabular}} & \textit{IEEE Xplore} & \cite{arcuri2020} \\ \hline

\textit{\begin{tabular}[c]{@{}l@{}}Testing Education: A Survey on a Global Scale\end{tabular}} & \textit{ACM DL} & \cite{melo2020} \\ \hline

\textit{\begin{tabular}[c]{@{}l@{}}The impact of Software Testing education on code\\ reliability: An empirical assessment\end{tabular}} & \textit{Scopus} & \cite{lemos2018} \\ \hline

\textit{\begin{tabular}[c]{@{}l@{}}Towards a Conversational Agent to Support the\\ Software Testing Education\end{tabular}} & \textit{ACM DL} & \cite{paschoal2019}\\ \hline

\textit{\begin{tabular}[c]{@{}l@{}}Use of JiTT in a Graduate Software Testing Course:\\ An Experience Report\end{tabular}} & \textit{IEEE Xplore} & \cite{martinez2018} \\ \hline

\end{tabular}
\end{table}

Visando obter as informações necessárias para se atender ao intuito do MSL, 11 itens foram estabelecidos para coleta e extração de dados. Este processo foi efetuado por intermédio da leitura completa dos estudos selecionados, buscando catalogar as informações consideradas relevantes de cada estudo.

Os campos a seguir compõem o formulário de extração: ano de publicação, veículo de publicação, tipo de veículo de publicação, foco do ensino, tipo de pesquisa, método de pesquisa, metodologias de ensino, atividade de ER, atividade de TS, tipo da contribuição e \textit{soft skills}.

\section{Análise e Síntese} 
\label{sec:sintese}

Os dados analisados do formulário de extração permitem a resposta da QP1, sendo apresentada a análise e síntese do conjunto de estudos selecionados a seguir.

\vspace{0.3cm}
\noindent\textbf{QP1: Qual é o estado da arte sobre o ensino de ER e de TS?}
\vspace{0.3cm}

\begin{figure}[ht]
  \centering
  \includegraphics[width=\textwidth]{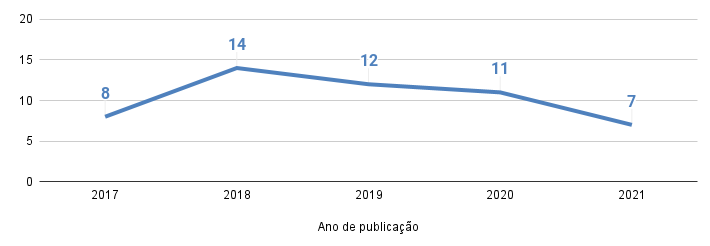}
  \caption{Número de artigos por ano de publicação.}
  \label{fig_ano_pub}
\end{figure}

 Ao todo, 52 artigos foram aceitos nesta pesquisa considerando os anos de 2017 a 2021. Em relação aos trabalhos aceitos, verificando-se por ano de publicação (Figura \ref{fig_ano_pub}), 2018 possuiu o maior índice de trabalhos, respectivamente, com 14 estudos -- representando um percentual de 26,9\% do total de publicações.

\begin{figure}[h]
  \centering
  \includegraphics[width=\textwidth]{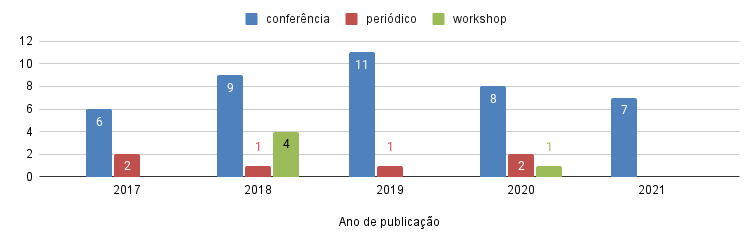}
  \caption{Veículos de publicação por ano.}
  \label{fig_tipo_veiculo}
\end{figure}

Por conseguinte, a grande parte dos artigos deriva de conferências, sendo o tipo principal de veículo de publicação retornado com 41 trabalhos (o que representou 78,8\%). Em sequência, os periódicos possuíram seis estudos (11,5\%) e por fim, cinco trabalhos foram advindos de \textit{workshops} (9,6\%). Em uma análise temporal, a  Figura \ref{fig_tipo_veiculo} demonstra que no ano de 2021 só foram aceitos estudos publicados em conferências, e desde 2017 possuiu majoritariamente o maior número de artigos. Acredita-se que esta maior prevalência de conferências ocorra devido ao fato que são consideradas os veículos com informações mais atuais dado o tempo de submissão e avaliação diminuto (se comparado aos periódicos), o que facilita a disseminação de práticas de ensino em voga pela comunidade da área de Computação.

Nesse sentido, foram encontrados 27 veículos de publicação distintos entre si. A conferência \textit{Brazilian Symposium on Software Engineering} (SBES) foi a que possuiu maior número de trabalhos, com seis estudos -- ao passo que também representa o protagonismo de pesquisas brasileiras no cenário de educação em ES. Posteriormente, o segundo veículo com mais trabalhos foi o \textit{IEEE Frontiers in Education Conference} (FIE), com cinco publicações aceitas.

Quanto ao foco do ensino, buscou-se verificar o alinhamento entre ER e TS, de forma integrada. Todavia, nenhum dos estudos apresentou de maneira explícita este relacionamento entre as áreas, focando sempre nestas atividades do ciclo de vida do software individualmente. Sendo assim, 27 dos artigos enfocaram em TS (51,9\%), enquanto 25 trabalhos destacaram o ensino em ER (48,1\%). É importante pontuar que apesar da literatura trazer indícios da necessidade de alinhamento entre ER e TS \cite{barmi2011,bjarnason2014}, \textit{as pesquisas recuperadas demonstram evidências de que o que está sendo ensinado na academia não apresenta aos acadêmicos a ligação destas duas etapas, comprometendo esta visão integrada do futuro profissional}.

\begin{figure}[h]
  \centering
  \includegraphics[width=\textwidth]{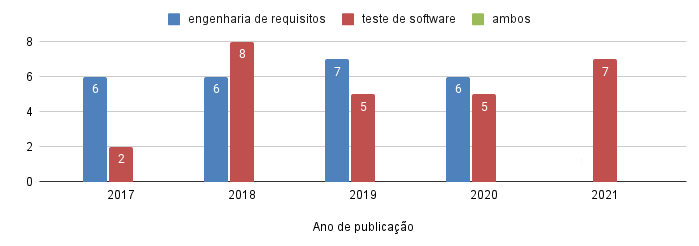}
  \caption{Foco do ensino por ano de publicação.}
  \label{fig_foco_ensino}
\end{figure}

Analisando o foco de ensino em relação ao ano de publicação (Figura \ref{fig_foco_ensino}), os anos de 2017, 2019 e 2020 possuíram mais artigos com enfoque no ensino de ER. Já nos anos de 2018 e 2021, TS se destacou, sendo que no ano mais recente de análise (2021), apenas pesquisas concentradas no tema de educação em TS foram aceitas como relevantes neste MSL. Como destacado pelo trabalho de Garousi et al. \cite{garousi2020msl}, este tópico está se tornando mais ativo, \textit{devido à elevada demanda por profissionais qualificados em TS, o que requer que estes conteúdos estejam inseridos dentro de cursos universitários}.

Acerca das metodologias de ensino adotadas, de forma geral a mais citada tratou-se de \textit{role-playing}, que foi encontrada em 10 estudos (12,8\%), a qual objetiva a simulação/inversão de papéis \cite{svensson2017}, a exemplo de um estudante que simula seu papel como engenheiro de requisitos, outrora como cliente, etc. Em segundo, a aprendizagem baseada em problemas esteve presente em nove artigos (11,5\%). Ressalta-se que vários trabalhos adotaram mais de uma metodologia de ensino, sendo este número uma contagem geral de quantas vezes essa metodologia foi praticada nos diferentes estudos. A classificação das metodologias foi realizada, em suma, segundo o recurso educacional aberto produzido por \cite{mendes2018}. A Figura \ref{fig_metodologias} apresenta a lista completa de metodologias e a quantidade de citações nos estudos. 

Ao avaliar as metodologias de ensino no que tange à requisitos e testes, \textit{role-playing} aparece exclusivamente na atividade de ER, sendo a mais proeminente com 10 estudos. Já em relação a TS, empataram com cinco estudos cada, a aprendizagem baseada em problemas e aula expositiva tradicional. Além disso, algumas metodologias ocorreram apenas em um determinado contexto, a exemplo de aprendizagem baseada em casos (com três estudos) que requer um conhecimento teórico prévio, presente apenas para TS. Ao passo que a aprendizagem baseada em competências (também com três estudos) esteve presente apenas no ensino de ER, como detalhado na Figura \ref{fig_metod_foco}. 

\begin{landscape}
\centering
\begin{figure}[t!]
  \centering
  \includegraphics[width=24cm]{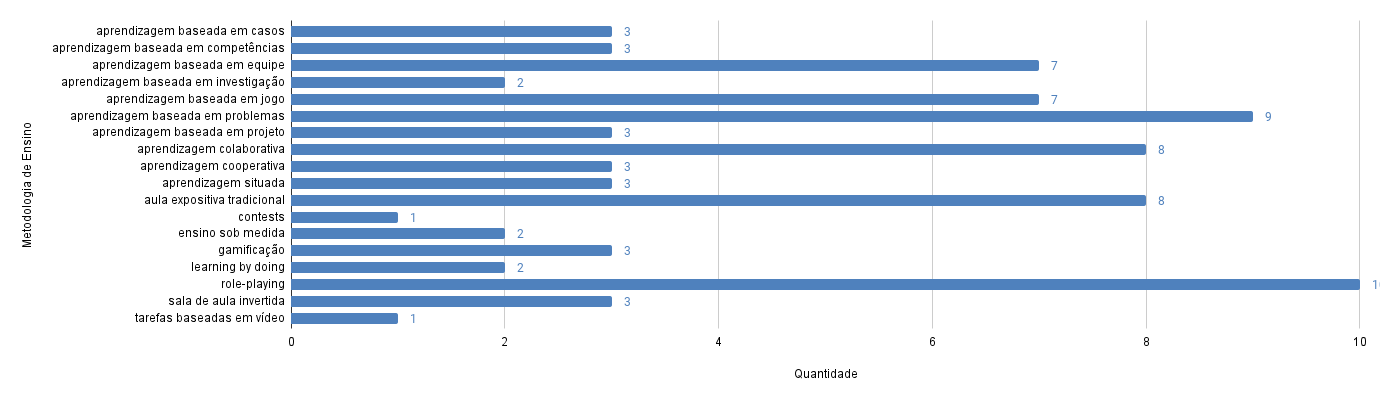}
  \caption{Metodologias de ensino presentes nos estudos aceitos.}
  \label{fig_metodologias}
\end{figure}

\begin{figure}[b!]
  \centering
  \includegraphics[width=24cm]{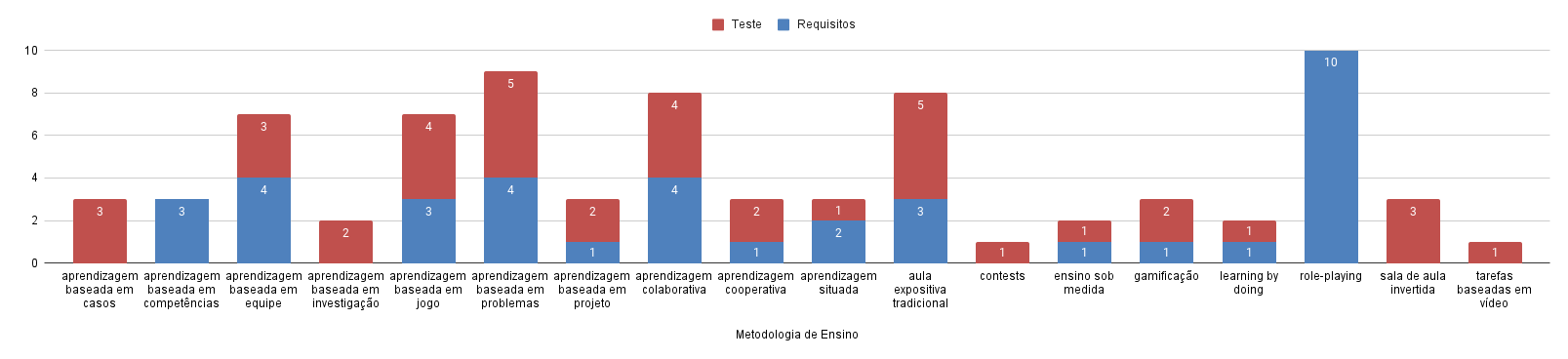}
  \caption{Metodologias de ensino com enfoque em requisitos ou testes.}
  \label{fig_metod_foco}
\end{figure}

\end{landscape}

\begin{figure}[h]
  \centering
  \includegraphics[width=\textwidth]{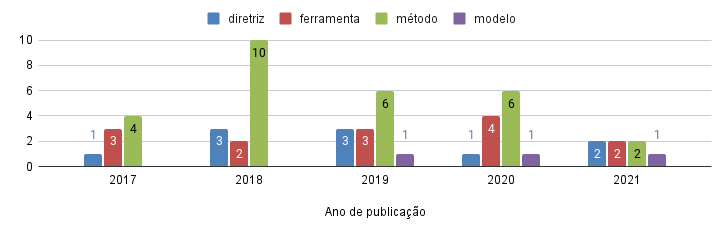}
  \caption{Tipo da contribuição por ano de publicação.}
  \label{fig_contrib}
\end{figure}

Quanto ao tipo de contribuição, quatro tipos principais foram verificados nos estudos, a saber: diretriz, ferramenta, método e modelo. Esta classificação segue os estudos de \cite{barmi2011, ouhbi2015, petersen2015}. No contexto geral, \textit{método} foi o tipo mais retornado pelos artigos analisados, com 28 trabalhos (50,9\%). Por ano, esta também foi a contribuição mais proeminente apresentada pelas publicações (Figura \ref{fig_contrib}).

Desta forma, é perceptível a preocupação da comunidade científica em propor maneiras de se ensinar tópicos da ES, neste caso, requisitos e teste (mesmo que os estudos ainda não apontem a integração de ambos). Portanto, \textit{a necessidade de estratégias pedagógicas representa um espaço ainda não consolidado e passível de desenvolvimento de pesquisas, ao passo que a própria forma de ensino também é algo mutável}. Ou seja, existe a necessidade de se adaptar à realidade dos estudantes, ou mesmo, à evolução da própria área de conhecimento, que precisa estar ciente das demandas da indústria.

Verificando-se o tipo de pesquisa, de acordo com a classificação de \cite{petersen2015}, três tipos foram considerados quando da avaliação empírica: relato de experiência, pesquisa de avaliação e pesquisa de validação. Em relação aos dois últimos, a pesquisa de validação diferencia-se por não ser aquela efetuada diretamente na prática, ou seja, é adotada geralmente com estudantes na academia. O contrário disso é a pesquisa de avaliação, que ocorre em um contexto industrial realístico envolvendo profissionais.

\begin{figure}[h]
  \centering
  \includegraphics[width=\textwidth]{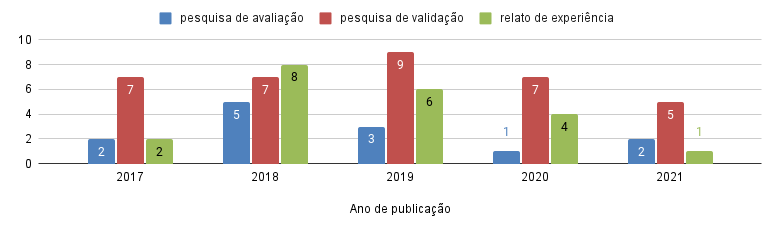}
  \caption{Tipo de pesquisa por ano de publicação.}
  \label{fig_pesquisa}
\end{figure}

Assim, a maioria dos estudos possuiu pesquisas com estudantes no formato de validação. Isso foi analisado em 35 artigos, o que representa 50,7\%. Em seguida, destacaram-se relatos de experiência com 18 estudos (30,4\%) e por fim, pesquisas de avaliação com profissionais -- com 13 trabalhos (18,8\%). A Figura \ref{fig_pesquisa} traz essa análise ano a ano, sendo que apenas em 2018 a pesquisa de validação não possuiu o maior índice de publicações. \textit{Acredita-se que a predominância de pesquisas de validação seja inerente à dificuldade da academia em estabelecer contato com a própria indústria e vice-versa, demonstrando uma lacuna entre os dois setores}. Alguns estudos destacam a dificuldade inclusive de adotar projetos práticos derivados da indústria, devido a questões de tempo, logística, proteção de dados, dentre outros fatores \cite{andrade2019}. Outra característica é a disponibilidade de profissionais da indústria para participar de experimentações para avaliação das pesquisas advindas da academia.

No que concerne às atividades de ER \cite{pressman2016}, há estudos que abarcam todas as atividades do processo de ER, desde elicitação, análise, especificação, validação e gerenciamento de requisitos, ou ainda, existem trabalhos que enfocam somente em uma delas, p.ex, elicitação ou análise. Nesse cenário, o maior contigente de estudos se preocuparam com o ensino de elicitação (32,4\%), enquanto, em menor número, gerenciamento de requisitos atingiu apenas 7,4\%. Já quanto às atividades de TS \cite{swebok2014}, a mais citada foi geração de casos de testes, com 20,7\%, enquanto, a menos levantada, tratou-se de rastreamento de defeitos, com 9,5\%.

\begin{figure}[h]
  \centering
  \includegraphics[width=\textwidth]{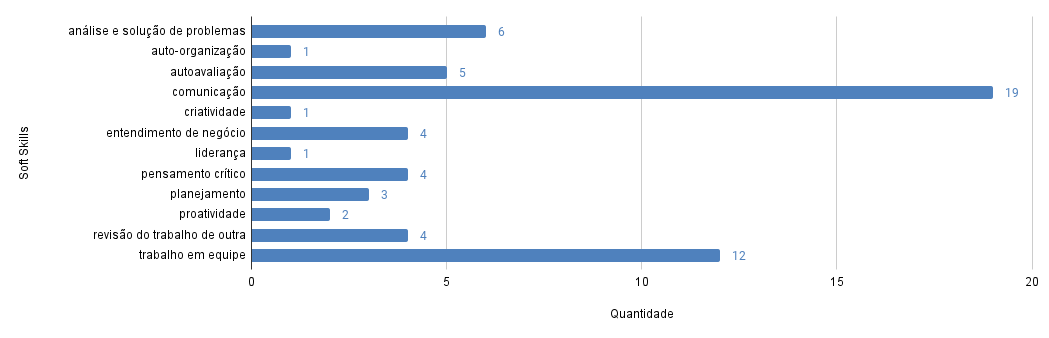}
  \caption{\textit{Soft skills} presentes nos estudos aceitos.}
  \label{fig_ss}
\end{figure}

Por conseguinte, também foram recuperadas nas publicações indícios do desenvolvimento de habilidades não técnicas, conhecidas como \textit{soft skills}. Foram buscadas por aquelas que visassem contribuir com atividades de ER ou TS. Estudos de \cite{garousi2020} elencaram \textit{soft skills} de trabalho em equipe, comunicação, liderança e pensamento crítico como de grande importância para um engenheiro de software. Além destas, outras \textit{soft skills} que foram citadas nos trabalhos também foram catalogadas. A Figura \ref{fig_ss} demonstra que boa parte dos estudos apontaram que comunicação foi a habilidade mais desenvolvida pelas atividades promovidas, presente em 19 trabalhos -- representando um percentual de 30,6\% do total de publicações. Este quantitativo considerou quando o artigo explicitamente apontou a promoção de \textit{soft skills} usando esta terminologia, ou ainda, mesmo que não declarasse abertamente ser uma \textit{soft skill}, mas que a forma de ensino possibilitou claramente sua promoção nos indivíduos.

Analisando as \textit{soft skills} relacionadas ao foco do ensino, a habilidade de comunicação foi predominante nas publicações de ER, presente em 15 artigos. Uma das hipóteses para isso se deve ao fato que \textit{a própria área de ER no seu processo elicitar, especificar, analisar e validar requisitos demanda bastante interação -- inclusive com os clientes e usuários finais, sendo esta \textit{skill} muito importante para a escrita de requisitos completos e sem ambiguidade}. Já para TS, o trabalho em equipe foi apontado como majoritário com cinco publicações. Exemplos de \textit{soft skills} como criatividade e liderança foram encontradas apenas em TS, mesmo que com um estudo cada (Figura \ref{fig_ss_foco}).

Para resumir os principais resultados, a Figura \ref{fig_bolhas} demonstra um gráfico de bolhas entre o foco de ensino, metodologias e \textit{soft skills} recuperadas nos estudos deste MSL. O tamanho da bolha é proporcional ao número de artigos presentes no par de categorias analisadas conforme cada um dos eixos do gráfico. Nesse cenário, \textit{role-playing} é a metodologia mais proeminente em ER, ao passo que, p. ex. aprendizagem baseada em projeto e gamificação foram citadas com apenas um estudo cada. Já comunicação é a \textit{soft skill} mais presente em ER, enquanto trabalho em equipe destaca-se em TS.

\section{Considerações Finais}
\label{sec:consid_msl}

Um estudo sistemático da literatura propicia uma visão geral de um determinado tópico de pesquisa, visando identificar tendências e lacunas. Sendo assim, dado os estudos relacionados ao ensino de ER e TS, é perceptível uma lacuna quanto ao alinhamento de requisitos e testes, já que ambas as atividades são ensinadas apenas de modo isolado. Outro fator é que mesmo que \textit{soft skills} sejam elementos importantes para a atuação profissional, este item não está sendo motivo de preocupação durante o exercício de estratégias de ensino, dado que um número significativo de estudos destaca somente a importância do aprendizado de conceitos técnicos de ER ou de TS, i.e, nem citam \textit{soft skills}.

Os achados do presente MSL, em especial sobre metodologias de ensino e \textit{soft skills}, podem ser cruzados com apontamentos das necessidades do mercado de software, quanto ao que se vem praticando na indústria atual. Nesse sentido, estudos de \cite{scatalon2019} apontam a existência de uma lacuna de conhecimento quanto à adoção de técnicas de histórias de usuário com cenários de teste. Essa lacuna negativa foi calculada mediante o que os profissionais atuantes na indústria brasileira aprenderam na graduação e o que eles realmente aplicam na carreira após formados. Corroborando para tal, \cite{benitti2017} também verificou que a grande maioria de empresas de software do país estão adotando histórias de usuário no desenvolvimento de projetos de software. Logo, é de suma importância que a academia possa apresentar formas de documentação em uso na indústria, em vistas de maior conhecimento e aprofundamento naquilo que o mercado carece \cite{epifanio2019}.

Com isso, como trabalhos futuros objetiva-se atualizar este MSL considerando um maior recorte temporal, idealmente, desde 2001, ano de publicação do Manifesto Ágil -- além de combinar demais estratégias de busca, como coleta manual e \textit{snowballing} dos estudos tidos como relevantes.

\begin{landscape}
\centering
\begin{figure}[]
  \centering
  \includegraphics[width=24cm]{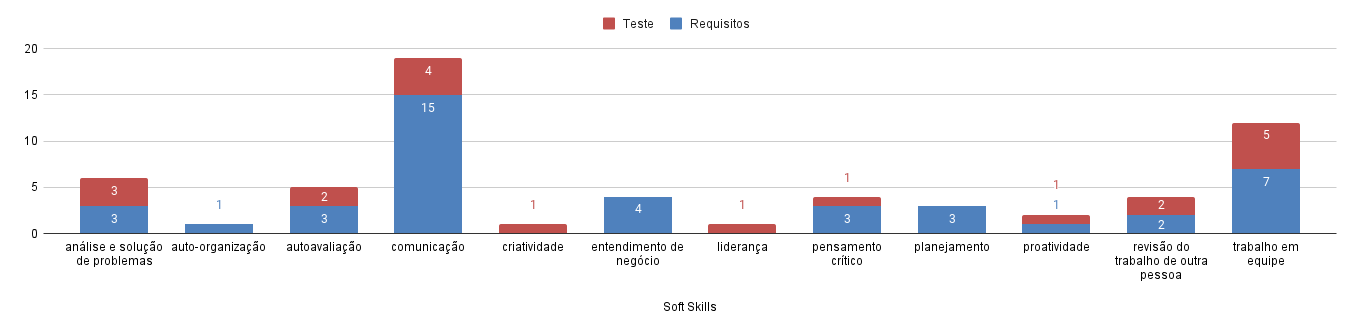}
  \caption{\textit{Soft skills} com enfoque em requisitos ou testes.}
  \label{fig_ss_foco}
\end{figure}

\begin{figure}[]
  \centering
  \includegraphics[width=24cm]{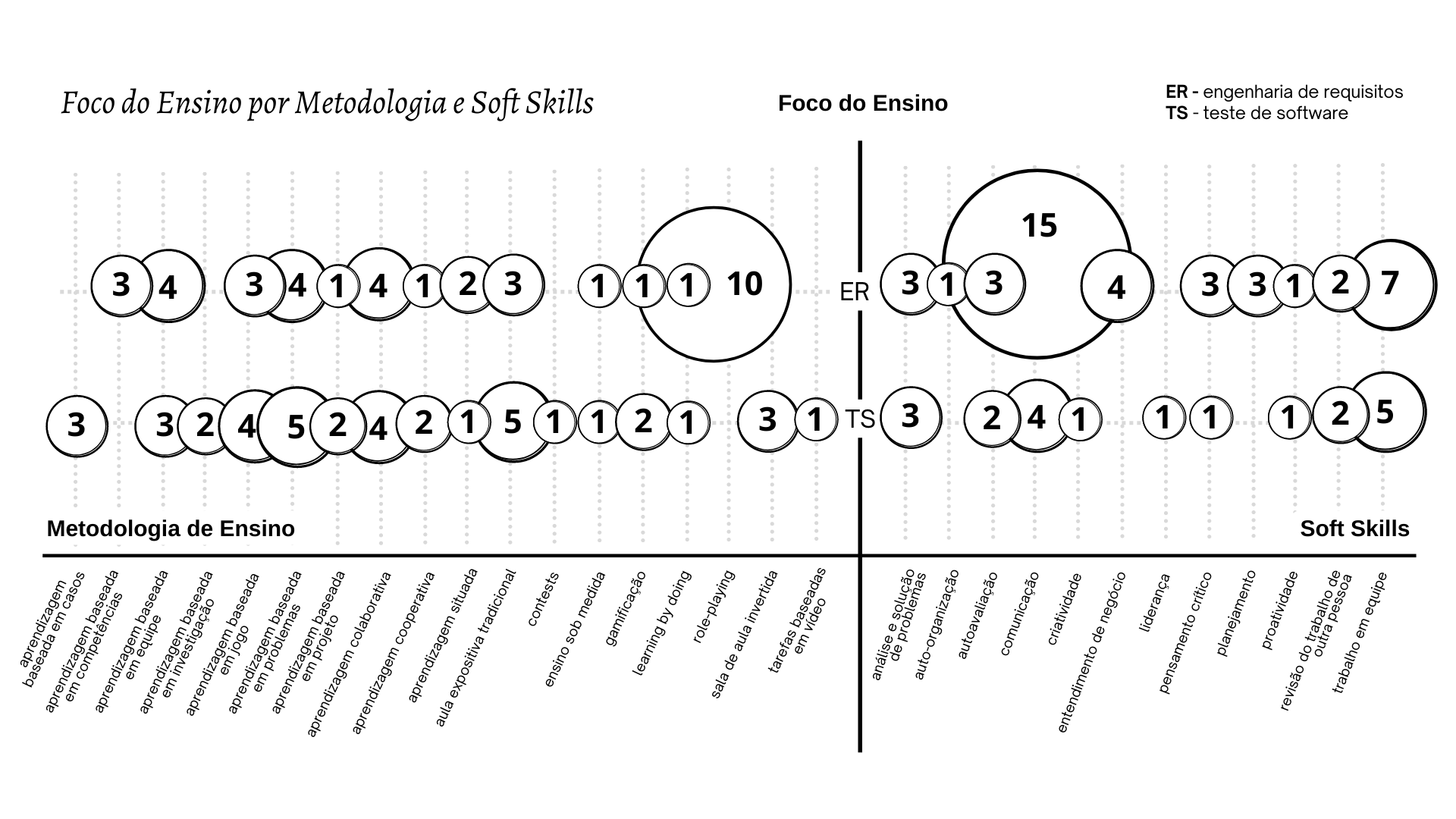}
  \caption{Gráfico de bolhas mapeando o alinhamento de ER e TS com metodologias de ensino e \textit{soft skills}.}
  \label{fig_bolhas}
\end{figure}

\end{landscape}

\bibliographystyle{sbc}
\bibliography{sbc-template}
\end{document}